\documentclass[conference]{IEEEtran}

\ifCLASSINFOpdf
\usepackage[pdftex]{graphicx}
\else
\fi

\usepackage{balance}
\usepackage{cite}
\usepackage{graphicx}
\usepackage{float}
\usepackage{amsmath}
\usepackage[utf8]{inputenc}
\usepackage[final]{pdfpages}
\usepackage{pdfpages}
\usepackage{lipsum}
\usepackage{textcase}
\usepackage{url}
\usepackage{amsmath,esint}
\usepackage{epstopdf}
\usepackage{array}
\usepackage{color}
\usepackage{subcaption}

\usepackage{bm}

\DeclareRobustCommand{\uvec}[1]{{%
  \ifcsname uvec#1\endcsname
     \csname uvec#1\endcsname
   \else
    \bm{\hat{\mathbf{#1}}}%
   \fi
}}

	\newcommand\blfootnote[1]{%
		\begingroup
		\renewcommand\thefootnote{}\footnote{#1}%
		\addtocounter{footnote}{-1}%
		\endgroup
	}
\newcolumntype{P}[1]{>{\centering\arraybackslash}p{#1}}
\newcolumntype{M}[1]{>{\centering\arraybackslash}m{#1}}

\pdfoutput=1

\begin{document}
	
\title{Coverage Enhancement for mmWave Communications using Passive Reflectors}

\author{
\IEEEauthorblockN{Wahab Khawaja\IEEEauthorrefmark{1}, Ozgur Ozdemir\IEEEauthorrefmark{1}, Yavuz Yapici\IEEEauthorrefmark{1}, Ismail Guvenc\IEEEauthorrefmark{1}, and Yuichi Kakishima\IEEEauthorrefmark{2} 
}
\IEEEauthorblockA{\IEEEauthorrefmark{1}Department of Electrical and Computer Engineering, North Carolina State University, Raleigh, NC}
\IEEEauthorblockA{\IEEEauthorrefmark{2}DOCOMO Innovations, Inc., Palo Alto, CA}
Email: \{wkhawaj, oozdemi, yyapici, iguvenc\}@ncsu.edu, kakishima@docomoinnovations.com
	
}

\maketitle

\blfootnote{This work has been supported in part by NASA under the Federal Award ID number
NNX17AJ94A and by DOCOMO Innovations, Inc.}

\begin{abstract}
Millimeter wave (mmWave) technology is expected to dominate the future 5G networks mainly due to large spectrum available at these frequencies. However, coverage deteriorates significantly at mmWave frequencies due to higher path loss, especially for the non-line-of-sight~(NLOS) scenarios.  
In this work, we explore the use of passive reflectors for improving mmWave signal coverage in NLOS indoor areas.  
Measurements are carried out using the PXI-based mmWave transceiver platforms from National Instruments operating at $28$~GHz, and the results are compared with the outcomes of ray tracing (RT) simulations in a similar environment. 
For both the measurements and ray tracing simulations, different shapes of metallic passive reflectors are used to observe the coverage (signal strength) statistics on a receiver grid in an NLOS area. For a square metallic sheet reflector of size $24\times24$ in$^2$ and $33\times33$ in$^2$, we observe a significant increase in the received power in the NLOS region, with a median gain of $20$~dB when compared to no reflector case. The cylindrical reflector shows more uniform coverage on the receiver grid as compared to flat reflectors that are more directional.


\end{abstract}

\IEEEpeerreviewmaketitle

\section{Introduction}
There is an ever increasing demand for higher communication data rates with newer applications requiring higher data bandwidths. The sub-$6$~GHz spectrum is reaching its limits due to spectrum congestion. With the  opening of mmWave spectrum by FCC~\cite{FCC_28G}, researchers have been exploring the realization of $5$G communication networks at mmWave frequencies. 
 A major bottleneck for mmWave propagation in the free space is high attenuation that makes radio frequency planning in the non-line-of-sight~(NLOS) very difficult~\cite{haneda20165g,rupasinghesystem}. Solutions to this problem may include use of high transmit power, high sensitivity receivers, and deployment of multiple access points or repeaters to improve link quality. However, increasing the transmit power and receiver sensitivity beyond a given limit may not be practical due to sophisticated and expensive equipment required. Similarly, introducing multiple access points and repeaters is not economically feasible.

A practical solution for improving mmWave propagation in NLOS areas can be the use of passive metallic reflectors. The reflection properties are especially better at higher frequencies due to smaller skin depth \cite{reflection}. This solution can be attractive due to large life span, low maintenance cost, ease in interoperability, and low initial investment costs when compared to repeaters, consisting of active elements. 
Passive metallic reflectors have been used in the past for microwave links for long distance communications such as satellite communications~\cite{NASA_refl,Literature4,Literature5}, and base station to base station microwave links~\cite{Microwave_refl}. However, there are limited studies available in the literature on the use of passive reflectors for communication with user equipment (see e.g.~\cite{Literature6,Literature7}), primarily due to their lower efficiency as compared to active repeaters. 

\begin{figure}[!t]
	\centering
	\includegraphics[width=0.95\columnwidth]{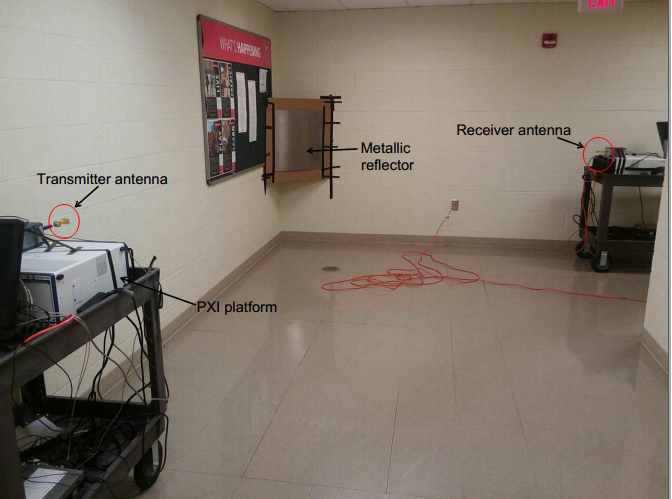}
	\caption{Measurement scenario in the basement corridor of Engineering Building~II at North Carolina State University for flat square sheet aluminum reflector of size $24\times24$~in$^2$, at an azimuth angle $\theta = 45^{\circ}$.}\label{Fig:Reflector_scenario}
\end{figure}

Passive reflectors can prove to be more useful for improving coverage at mmWave frequencies (see Fig.~\ref{Fig:Reflector_scenario}), due to better reflection properties in those frequencies. However, similar to low frequency microwave communications, 
there are limited studies available to date in the literature on the use of passive reflectors for mmWave communications. In \cite{Literature1}, a parabolic reflector is introduced behind a patch antenna for a hand held device operating at $60$~GHz. The introduction of parabolic reflector helps to counter the shadowing introduced by finger while operating the device. Simulations were carried out indicating a gain of $19$~dB - $25$~dB. In \cite{Literature2}, wideband channel sounding measurements at $60$~GHz were carried out to evaluate the reflecting properties of different building materials in indoor and outdoor environments. In~\cite{Literature3}, a parabolic passive reflector is used in outdoors at mmWave frequencies to reflect the low energy signal of NLOS path to shadowed zones. Numerical results were used to indicate that there is significant increase in the coverage area by using multiple reflectors.

In this work, as illustrated in Fig.~\ref{Fig:Reflector_scenario}, we have performed indoor measurements using different size and shape 
passive aluminum reflectors at $28$~GHz in an NLOS scenario using National Instruments~(NI) mmWave 
PXI platform with directional horn antennas. The received power is observed to improve 
for all the reflector shapes when compared to no reflector case. The received power for flat square sheet reflectors is observed to be higher when compared to cylindrical and spherical shaped reflectors. With $24\times24$ in$^2$ and $33\times33$ in$^2$ flat square sheet reflectors, we observe a median power gain of approximately $20$~dB along with better overall coverage on the receiver grid. Whereas for cylindrical reflector, we observe more uniform power distribution over the receiver grid.  
The measurement results are compared with the outcomes of ray tracing (RT) simulations incorporating the diffuse scattering phenomenon.             

\begin{figure}[!t]
	\centering
	\includegraphics[width=\columnwidth]{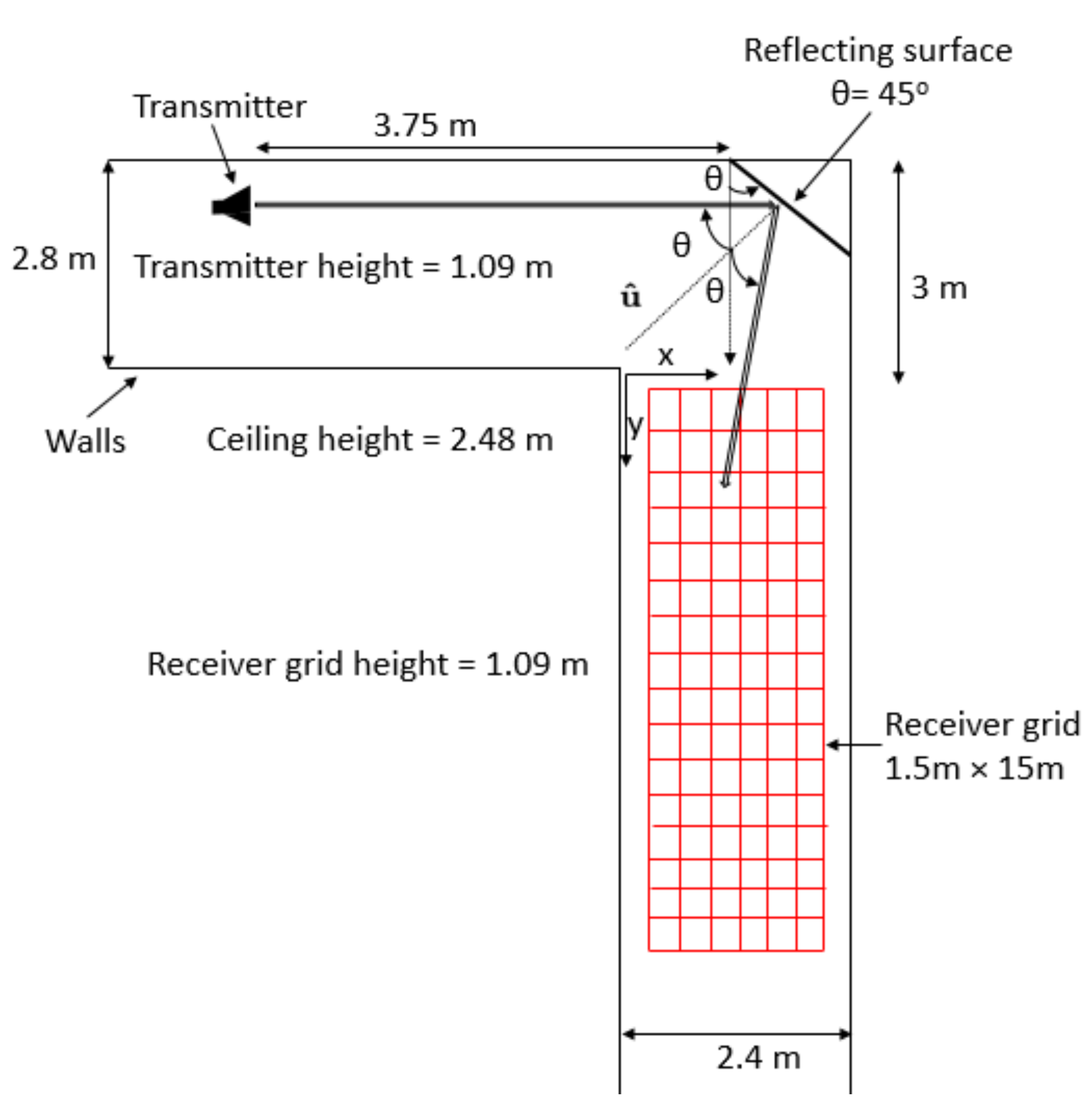}
	\caption{Geometrical model in the azimuth plane with a reflecting surface deployed at the corner of a corridor.}\label{Fig:Indoor_scenario_geometrical}
\end{figure}

\section{Propagation Measurements at 28 GHz}

\begin{figure}[!t]
	\centering
	\includegraphics[width=\columnwidth]{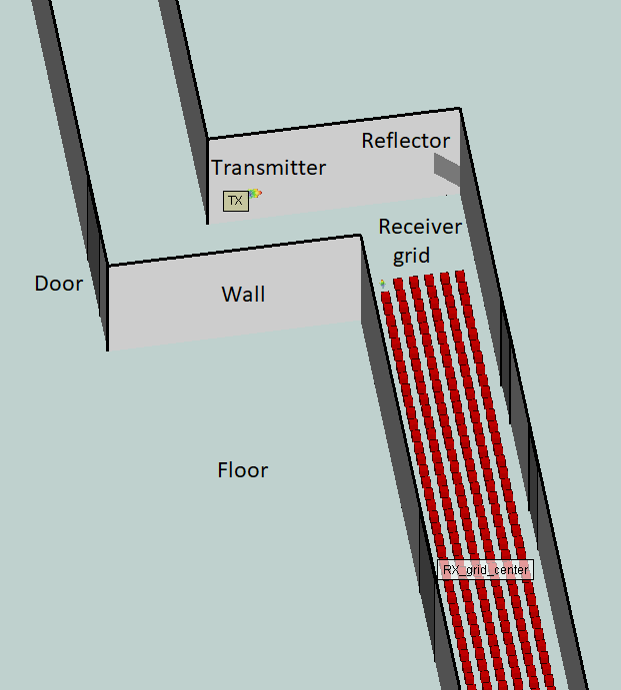}
	\caption{Indoor NLOS path scenario in the basement corridor of Engineering Building~II of North Carolina State University simulated in Wireless InSite for flat square sheet reflector $24\times24$~in$^2$ at an azimuth angle, $\theta = 45^{\circ}$.}\label{Fig:Indoor_scenario}
\end{figure}

Measurements were carried out in the basement corridor of Engineering Building~II at North Carolina State University as illustrated in Fig.~\ref{Fig:Reflector_scenario}. A more detailed geometrical sketch of the measurement environment is illustrated in Fig.~\ref{Fig:Indoor_scenario_geometrical}. The receiver is moved at different positions in the $(x,y)$ plane of the corridor to form a receiver grid. A similar geometry is generated using  the Remcom Wireless InSite ray tracing software to compare with the measurement outcomes, as shown in Fig.~\ref{Fig:Indoor_scenario} and will be explained in Section~\ref{Sec:RayTracing}.

\begin{figure}[!t]
	\centering
	\includegraphics[width=\columnwidth]{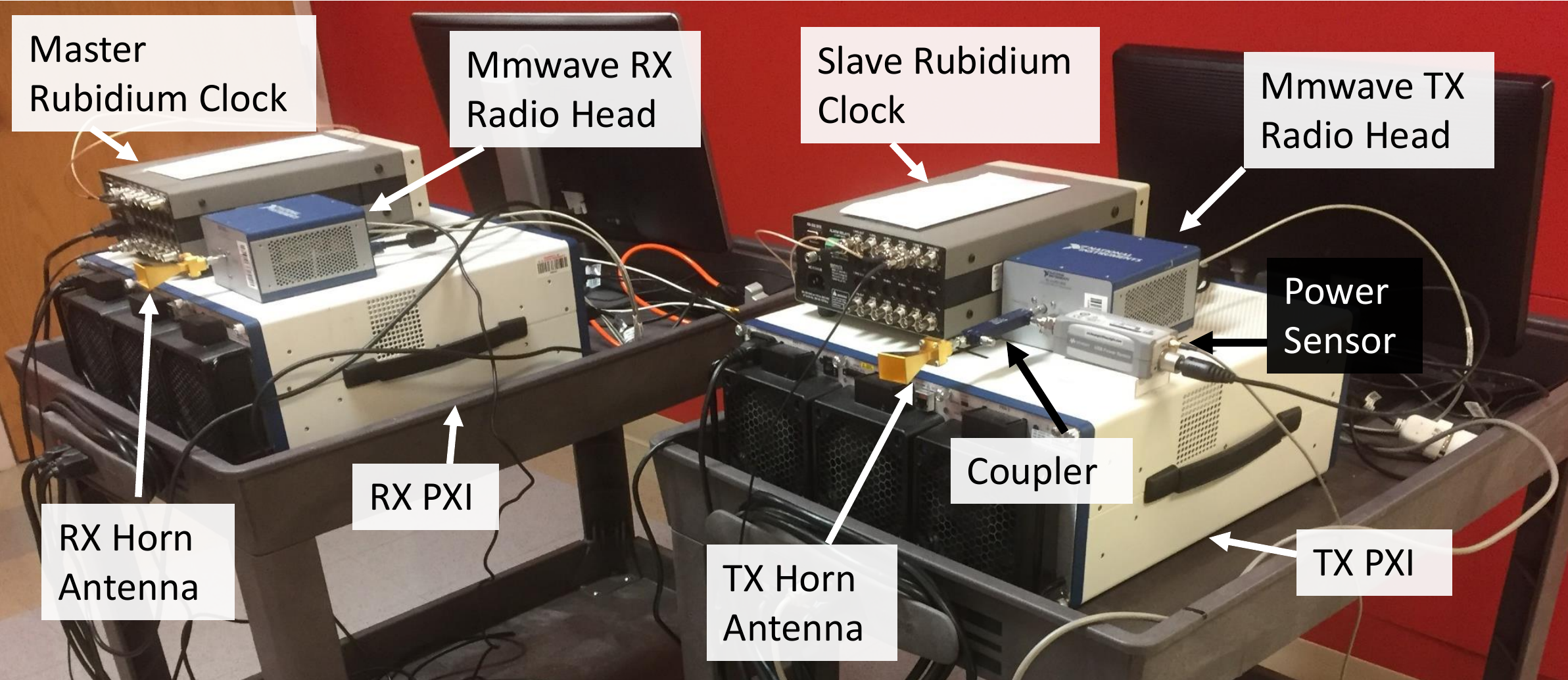}
	\caption{28 GHz channel sounder based on NI mmwave transceiver system.}\label{Fig:Setup}
\end{figure}

The measurements were performed using hardware based on NI mmWave transceiver system at 28 GHz~\cite{NImmwave} as shown in Fig.~\ref{Fig:Setup}. The system contains two PXI platforms: one transmitter platform and one receiver platform. Two rubidium (Rb) clocks, one master and one slave provides 10 MHz and PPS signals to the PXI timing and synchronization modules at the transmitter and the receiver. The master Rb clock trains the slave Rb clock so that clock signals are synchronized.{\footnote{The training needs to be performed before each measurement by connecting two Rb clocks by a coaxial cable. Once the trining is done then the cable can be disconnected and the systems can be separated without any cable connecting them.}} The coupler at the transmitter provides 30 dB attenuated signal for the power sensor to make power calibration. With power calibration we can convert the channel impulse response measurements from dB to dBm. A separate power calibration is performed at the receiver side as well.

The digital to analog converter at the transmitter PXI and the analog to digital converter at the receiver PXI have a high sampling rate of 3.072 GS/s. The channel sounder supports 1 GHz and 2 GHz modes of operation. The measurements for this paper are performed using the 2 GHz mode where the sounding signal duration is 1.33 $\mu$s, which also is the maximum measurable excess delay of the sounder. This mode provides a 0.65~ns delay resolution in the delay domain,  corresponding to 20~cm distance resolution. The analog to digital converter has around 60~dB dynamic range and this system can measure path loss up to 185~dB. 

In order get accurate channel measurements, 
we need to characterize the non-flat frequency response of the measurement hardware itself, and subsequently do a calibration to compensate for the impulse response due to the hardware. 
For calibration purposes a cable with fixed attenuators connects the transmitter to the receiver. Assuming the cable and the attenuators have flat response, the channel response of the hardware is measured. During actual measurements, the hardware response is equalized assuming hardware response does not vary over time. After this equalization we obtain the response of the actual over the air channel.

\begin{table}[!t]
	\begin{center}		
		\caption{Dimensions of reflectors used in the experiment.}\label{Table:Dimensions}
        	\begin{tabular}{|P{3.8cm}|P{3.8cm}|}
			\hline
			\textbf{Reflector Type}&\textbf{Dimensions of reflectors}\\
			\hline
           Sphere& $r = 13$~in \\
           \hline
          Cylinder& $r = 4.5$~in, $h = 18$~in   \\
            \hline
           Flat square sheet&$w=h= 33$~in, $24$~in, $12$~in \\
            \hline
           \end{tabular}
		\end{center}
\end{table}

To improve the coverage area in NLOS receiver region, we use reflectors of different shapes and sizes as illustrated in  Table~\ref{Table:Dimensions}, where $r$ represents the radius, while $w$, $h$ represent the width and height of the reflectors, respectively. The gain of a reflector at a given propagation path is dependent on the shape and cross-sectional area of the reflector, and  can be represented in terms of radar cross section~(RCS) that incorporates the shape and cross-sectional area~
\cite{RCS2}.  
The coverage can be improved in a given direction by changing the orientation of the reflector. If we want to steer the incident beam at an azimuth angle $2\theta$ to provide coverage to a particular area at a given azimuth angle, we need to tilt the reflector sheet in the azimuth plane by an angle $\theta$ as shown in Fig~\ref{Fig:Indoor_scenario}, where $\uvec u$ is the surface normal of the reflector sheet.


\section{Ray Tracing Simulations at 28 GHz}\label{Sec:RayTracing}

Simulations for the passive metallic reflectors at mmWave frequencies are performed using Remcom Wireless InSite ray tracing software, considering a similar indoor environment as shown in Fig.~\ref{Fig:Indoor_scenario}.  
A sinusoidal sounding signal at $28$~GHz is used, and the transmit power is set to $0$~dBm. Horn antennas \cite{Horn_antenna_sage} are used at both transmitter and the receiver grid. The antennas are vertically polarized with a gain of $17$~dBi. The antenna has $26^\circ$ and $24^\circ$ of half power beamwidth in the E-plane and H-plane, respectively. 
The simulation environment is similar to the actual environment with the inclusion of respective objects and their properties. 
The selection of wall, floor, ceiling, door and reflector materials are are made upon observing the real world materials in the measurement environment. \emph{ITU three layered drywall} is used for walls, ITU \emph{ceiling board} is used for ceilings, \emph{concrete} is used for floor, and a \emph{perfect conductor} is used for the door and the metallic reflector. All the materials are frequency sensitive at $28$~GHz. The dimensions of the simulation setup are same as in Fig.~\ref{Fig:Indoor_scenario_geometrical}. 

\begin{table}[!t]
	\begin{center}		
		\caption{Diffuse scattering parameters.}\label{Table:DS}
        	\begin{tabular}{|P{2.8cm}|P{3.5cm}|}
			\hline
			\textbf{Material}&\textbf{Diffuse scattering coefficient}\\
			\hline
            Perfect conductor& 0.1 \\
            \hline
            Concrete& 0.2   \\
            \hline
            Ceiling board& 0.25 \\
            \hline
            Layered dry wall& 0.3 \\
            \hline
           \end{tabular}
		\end{center}
 \end{table}

In addition to specular reflection at mmWave frequencies, diffuse scattering also occurs dominantly due to comparable size of wavelength of the transmitted wave and the dimensions of the irregularities of the surfaces that it encounters. In the simulations, diffuse scattering feature has been used to take into account this factor. The diffuse scattering model used in the simulations is directive model. Only the diffuse scattering coefficient is changed for different materials, whereas the other model parameters remain the same. Diffuse scattering coefficient of different materials used in the simulations are provided in Table~\ref{Table:DS}, where the materials with higher roughness have higher diffuse scattering coefficient.
The received power is obtained and summed coherently from the received MPCs at a given receiver location. This involves the phase of each MPC to be considered in the received power calculation. 


\section{Empirical and Simulation Results}
In this section, empirical and simulation results are analyzed for the indoor NLOS measurements with and without metallic reflectors. The received power is analyzed over a grid of dimensions $1.5{\rm m}\times15{\rm m}$.

\begin{figure}[!t]
	\begin{subfigure}{0.5\textwidth}
	\centering
	\includegraphics[width=\columnwidth]{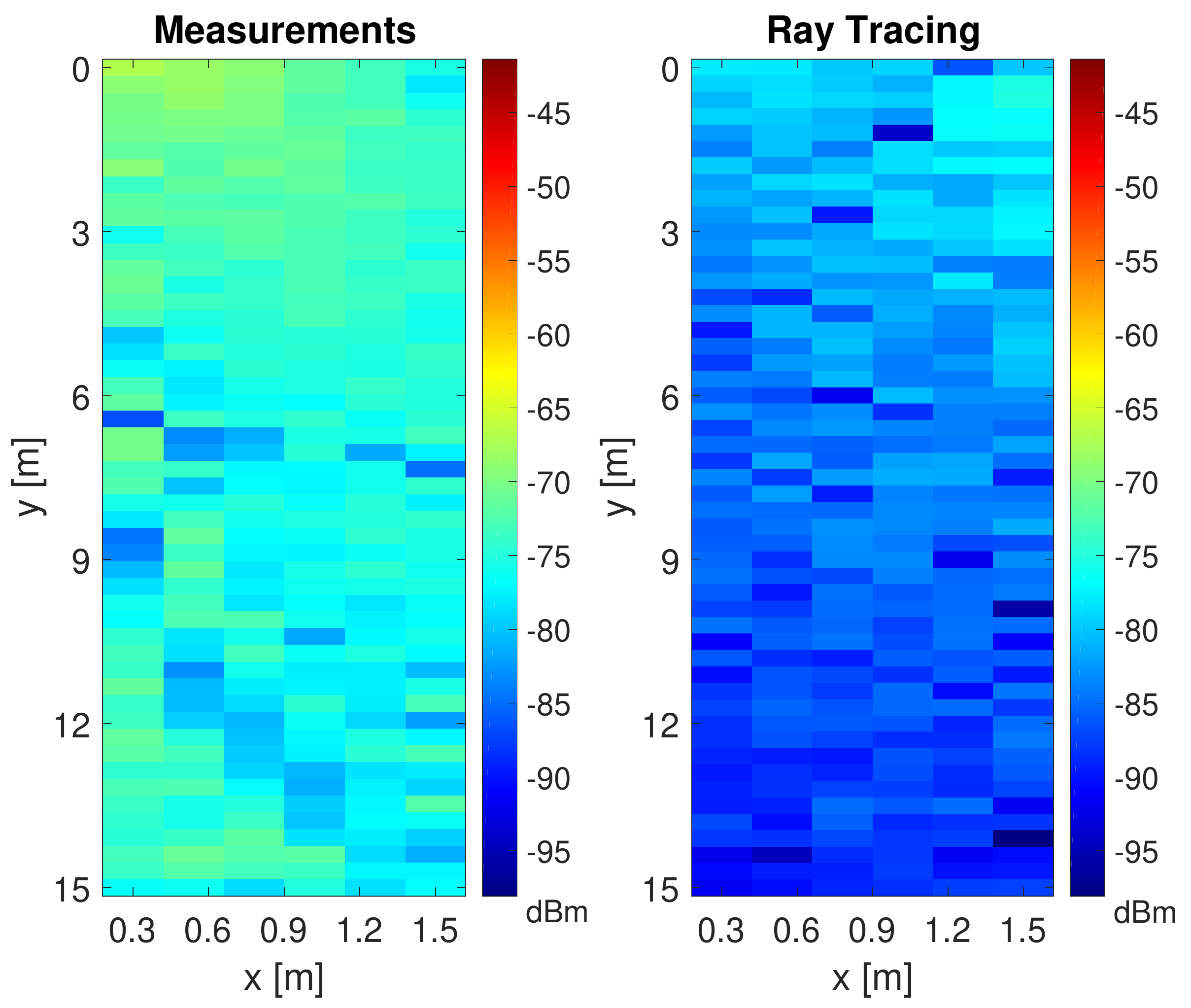}
	\caption{}
    \end{subfigure}			
	\begin{subfigure}{0.5\textwidth}
	\centering
    \includegraphics[width=\columnwidth]{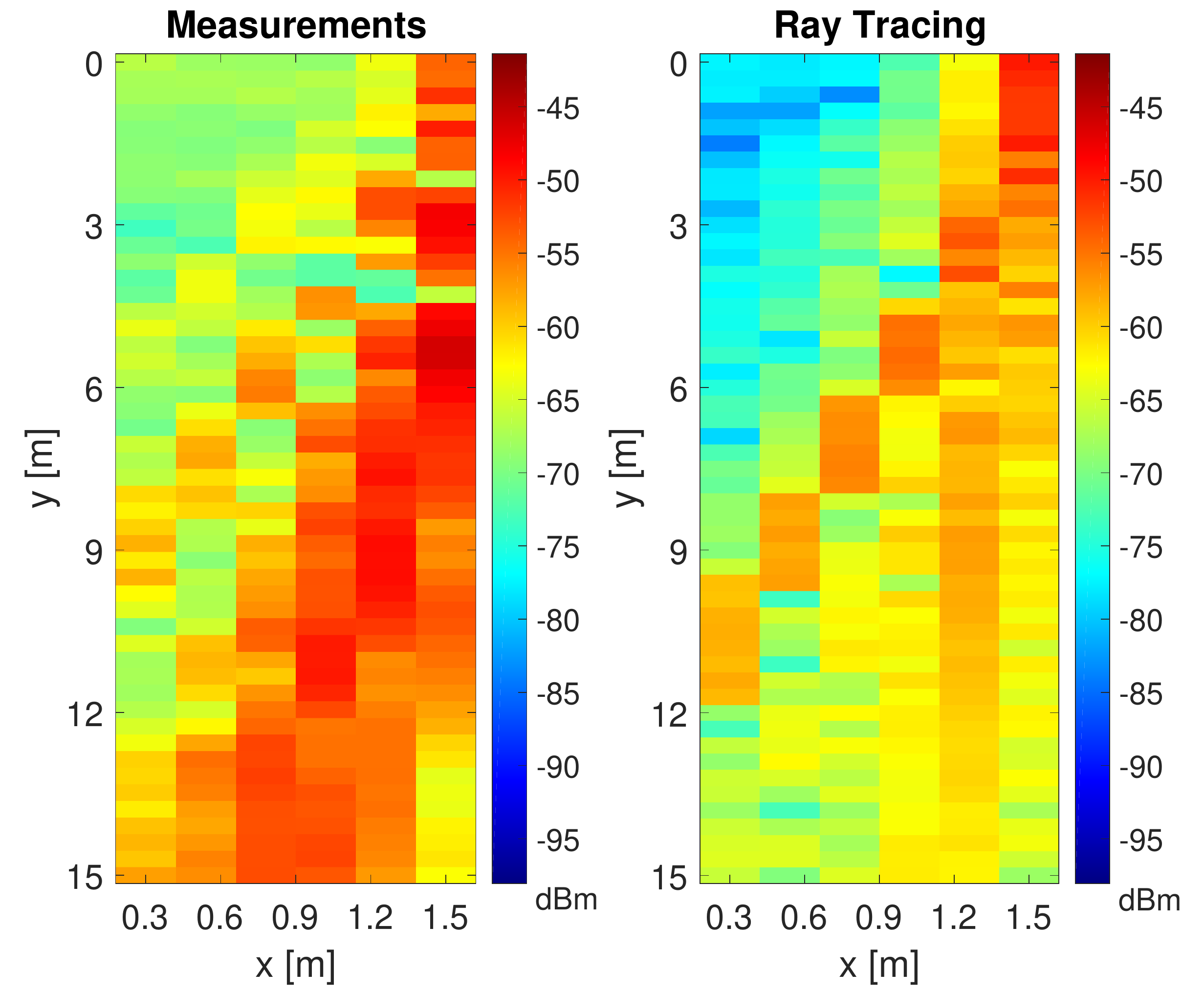}
	 \caption{}
     \end{subfigure}
    \caption{Received power results for (a) no reflector,  obtained using (left) measurements, and (right) ray tracing simulations; (b) $12\times12$~in$^2$ flat square aluminum sheet at $\theta = 45^\circ$,  obtained using (left) measurements, and (right) ray tracing simulations.}\label{Fig:No_refl_12inch}
\end{figure}

\subsection{Coverage with No Reflector}

In Fig.~\ref{Fig:No_refl_12inch}(a) (left), received power on the receive grid is shown using measurements when no reflector is utilized.   
On the other hand, Fig.~\ref{Fig:No_refl_12inch}(a) (right) shows the mmWave signal coverage using ray tracing simulations, again considering no reflector. We observe higher received power in case of measurements; possible effects contributing to this behavior can include  the additional scatterers in the real environment, and the specific values of  diffuse scattering coefficients and associated model parameters for simulations. 

\begin{figure}[!t]
	 	
	\begin{subfigure}{0.5\textwidth}
	\centering
    \includegraphics[width=\columnwidth]{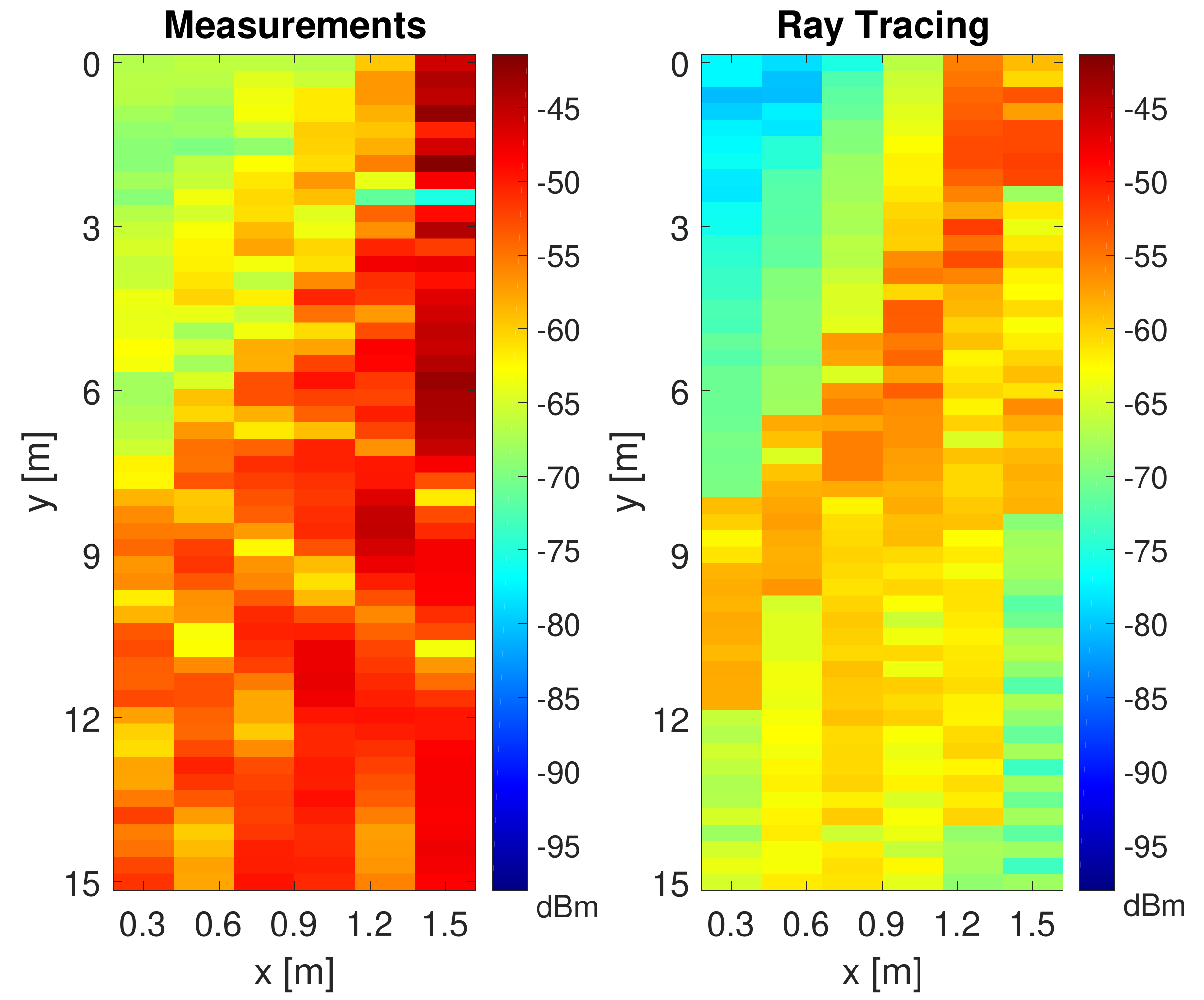}
	 \caption{}
     \end{subfigure}
     \begin{subfigure}{0.5\textwidth}
	\centering
    \includegraphics[width=\columnwidth]{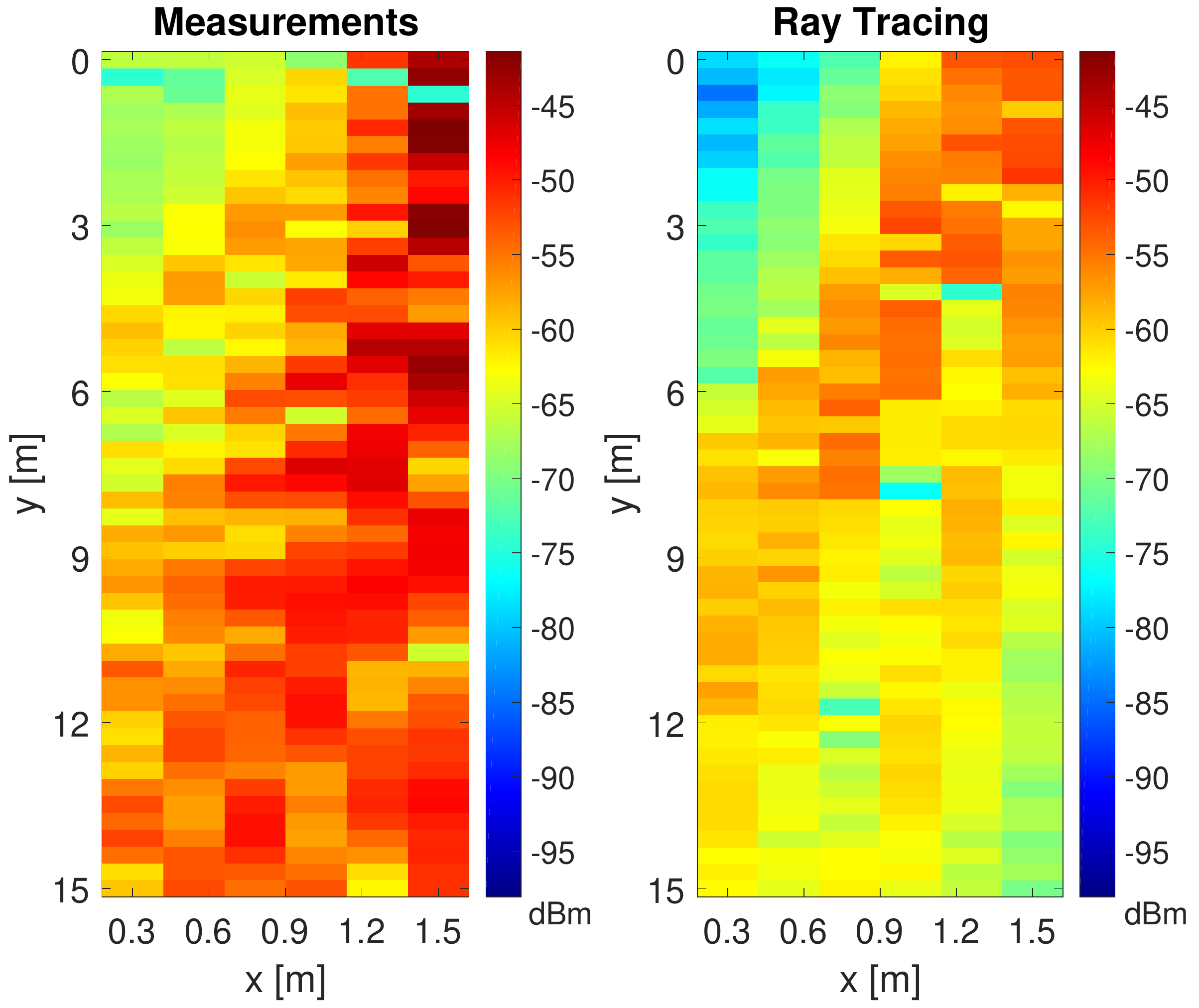}
	 \caption{}
     \end{subfigure}
    \caption{Received power results for (a) $24\times24$~in$^2$ flat square aluminum sheet at $\theta = 45^\circ$,  obtained using (left) measurements, and (right) ray tracing simulations; (b) $33\times33$~in$^2$ flat square aluminum sheet at $\theta = 45^\circ$,  obtained using (left) measurements, and (right) ray tracing simulations.}\label{Fig:flat_combine}
\end{figure}



\subsection{Coverage with Square Metal Reflectors}


The received power in case of $12\times12$~in$^2$ square sheet reflector oriented at an azimuth angle of $\theta = 45^{\circ}$ is shown in Fig.~\ref{Fig:No_refl_12inch}(b). We can observe a distinct directional pattern that enlarges with the distance over the grid. This directional pattern is perpendicular to the reflector, as the reflector is oriented at $45^\circ$. Moreover, we also observe second order reflections from the wall near the end of y-grid. In case of simulations, we observe lower reflected power as compared to measurements with similar power distribution over the grid. 

For the $24\times24$~in$^{2}$ square sheet reflector oriented at $45^\circ$, the received power falling on the grid is shown in Fig.~\ref{Fig:flat_combine}(a), where we observe the highest received power and large coverage in the azimuth plane on the receiver grid. The received power distribution over the grid is similar to as observed for $12\times12$~in$^{2}$ square sheet reflector. However, compared to the $12\times12$~in$^{2}$ reflector, we observe high power beam at larger azimuth angles on the grid, providing more coverage mainly due to large cross-section area of the reflector. The measurement/simulation results show close resemblance of power distribution, though in case of simulations the received power is small and more directed towards the wall.

The received power obtained over the grid for $33\times33$~in$^{2}$ reflector oriented at $45^\circ$ is shown in Fig.~\ref{Fig:flat_combine}(b). It can be observed that we have similar power distribution as observed for  $24\times24$~in$^{2}$ square sheet reflector. However, in comparison to $24\times24$~in$^{2}$ square sheet reflector, we observe large power at the start of the grid, but weakens slightly near the end of y-grid. Two reasons can be given for this behavior. First, at a given distance of transmitter from reflector, the intensity of the electric field is such that changing the size of the reflector will not play a significant role. Secondly, as the reflector and the receiver grid are at the same height, we may observe strong reflections from lower part of the reflector towards the ground and causing more destructive and constructive interference. 

\begin{figure}[!t]
	\begin{subfigure}{0.5\textwidth}
	\centering
    \includegraphics[width=\columnwidth]{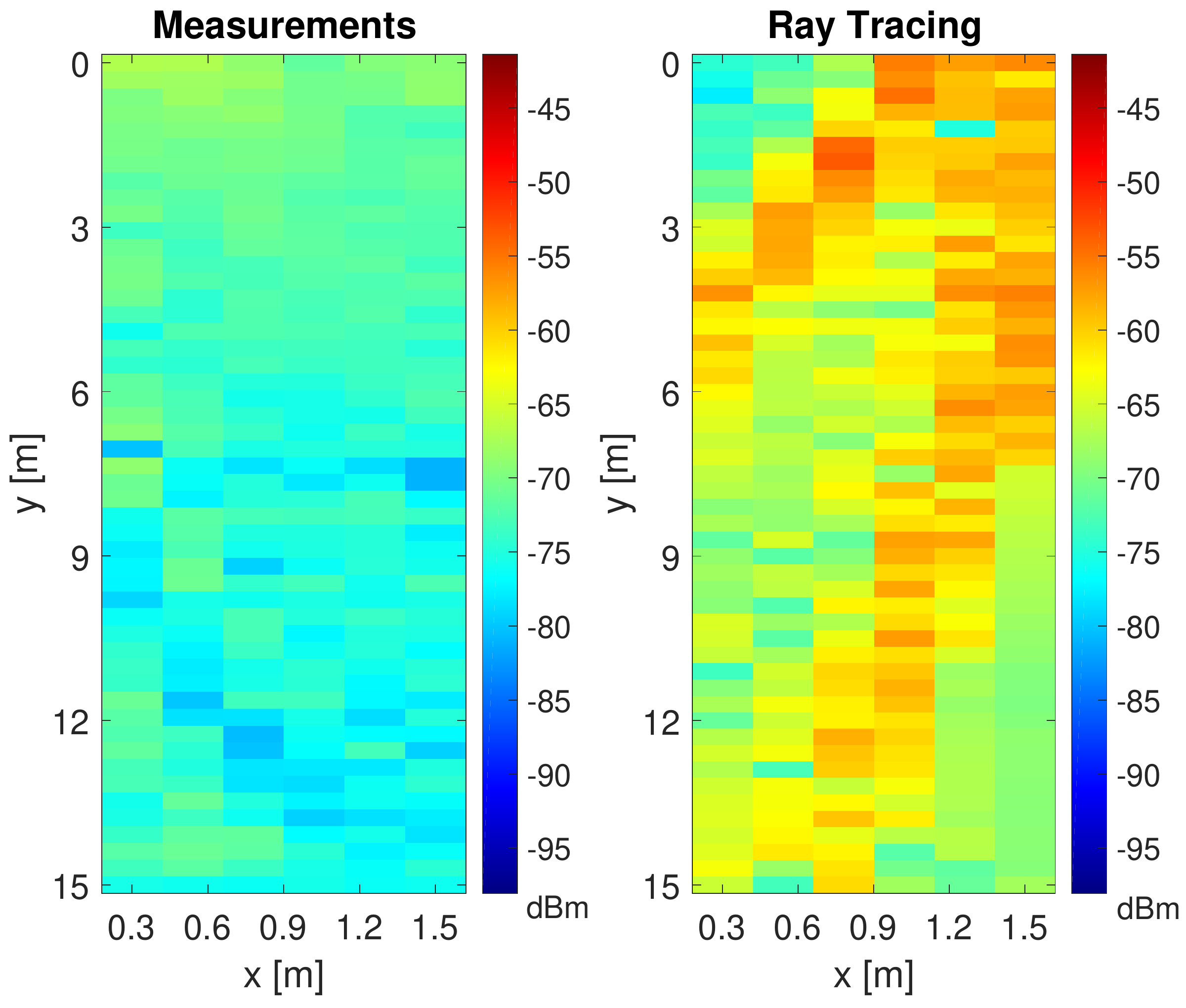}
	\caption{}
    \end{subfigure}
       
    \begin{subfigure}{0.5\textwidth}
	\centering           
	\includegraphics[width=\columnwidth]{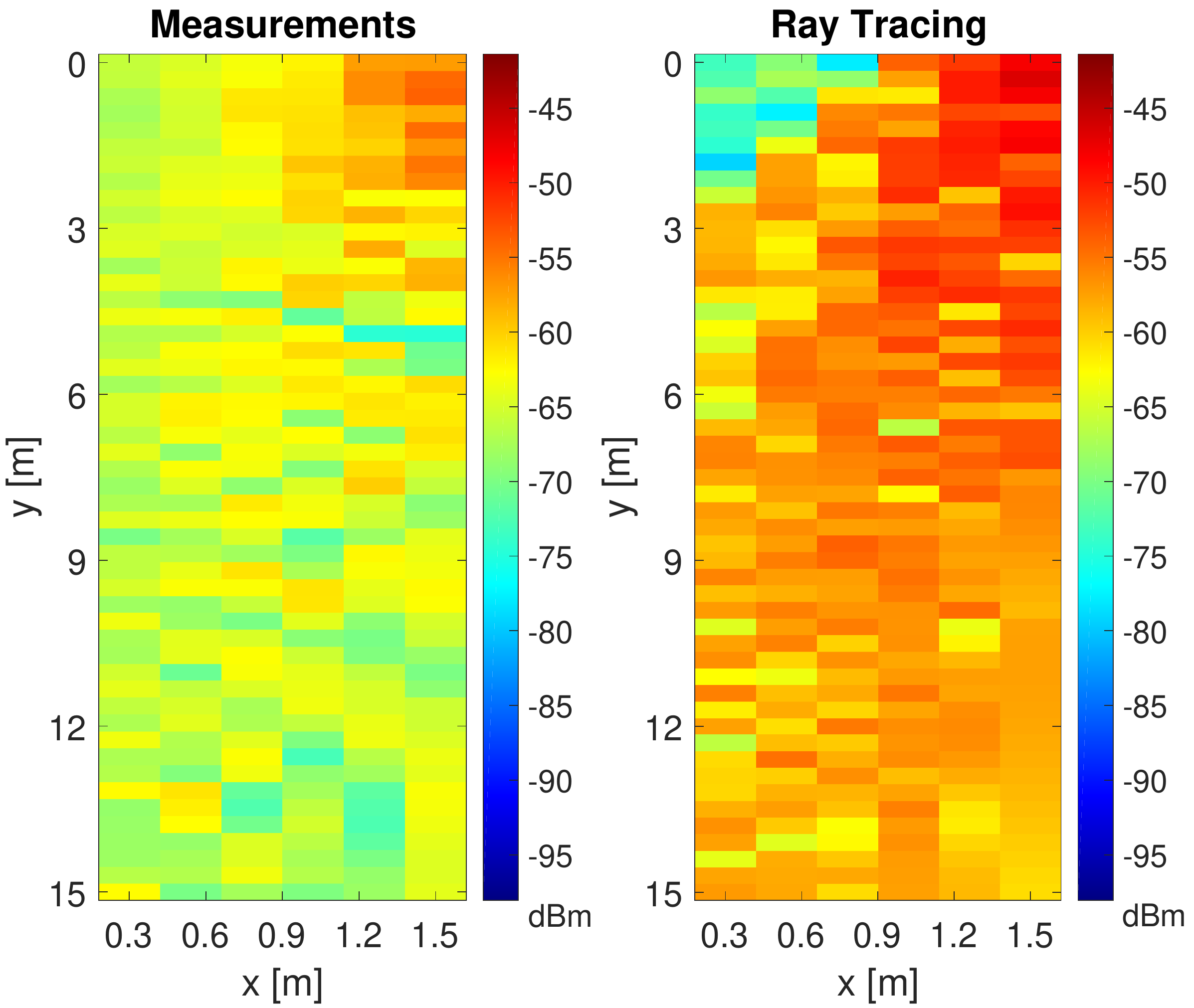}
	
    \caption{}
    \end{subfigure}
    \caption{Received power results for (a) metallic sphere obtained using (left) measurements, and (right) ray tracing simulations; (b) metallic cylinder obtained using (left) measurements, and (right) ray tracing simulations.}\label{Fig:sphere_cylinder_combine} 
\end{figure}



\subsection{Coverage with Spherical and Cylindrical  Metal Reflectors}

For the sphere reflector of radius $13$~in, the received power on the grid is approximately uniformly distributed at shorter distances as shown in Fig.~\ref{Fig:sphere_cylinder_combine}(a), proving that the gain of the spherical reflector is omni-directional. Similarly, for the cylindrical reflector of radius $4.5$~in and height $18$~in shown in Fig.~\ref{Fig:sphere_cylinder_combine}(b), the received power is more uniformly concentrated at shorter distance around the reflector similar to sphere due to circular curved shape of the reflector, diverging the incident beam in different directions. The received power with cylindrical case is higher as compared to sphere due to larger effective area exposed to the incident beam. This can be validated from the simulation results in Fig.~\ref{Fig:sphere_cylinder_combine}~(right).

In all experiments, the total cross-sectional area of the $24\times24$~in$^2$ flat sheet reflector, the sphere, and the cylinder were the same. However,  we observed higher received power in case of $24\times24$~in$^2$ square flat sheet, due to the larger cross section area exposed to the incident beam when compared to cylinder and sphere.

\begin{figure}[!t]
	\begin{subfigure}{0.5\textwidth}
	\centering
    \includegraphics[width=0.97\columnwidth]{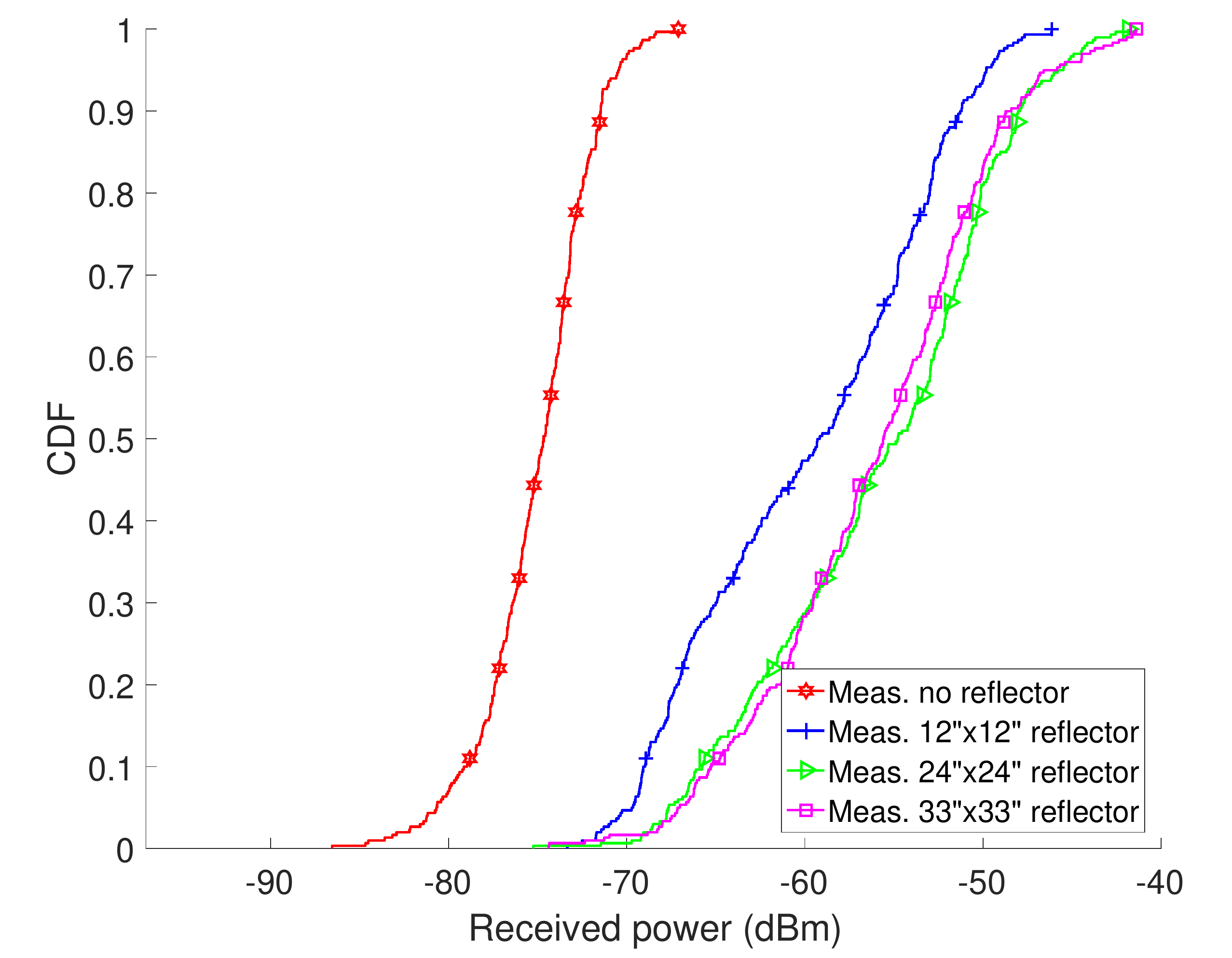}
    \caption{}
    \end{subfigure}	

 	\begin{subfigure}{0.5\textwidth}
    \centering
	\includegraphics[width=0.97\columnwidth]{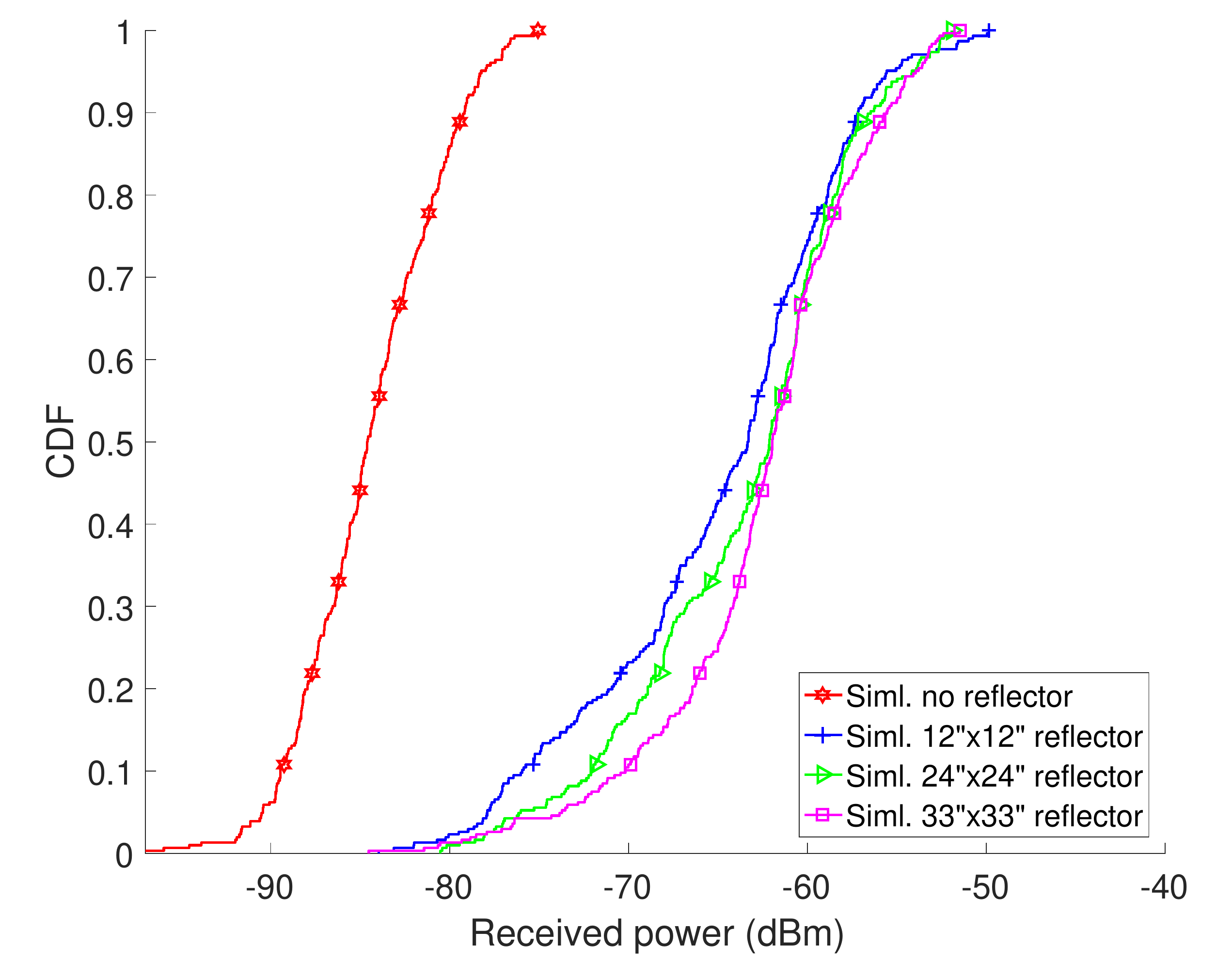}
	  \caption{}  
    \end{subfigure}
    \caption{CDF of received power for no reflector, $12\times12$~in$^2$, $24\times24$~in$^2$, and $33\times33$~in$^2$ flat square sheet reflectors from (a) measurements, (b) simulations.}\label{Fig:CDF_received_power_flat_sheets}
\end{figure}


\subsection{CDF of Received Power with/without Reflector}

The cumulative distribution function~(CDF) plots of received power for flat sheet reflectors and no reflector are shown in Fig.~\ref{Fig:CDF_received_power_flat_sheets}. For no reflector, we have lower received power and lower variance of the received power over the receiver grid. The $24\times24$~in$^2$ and $33\times33$~in$^2$ flat square sheet reflectors have similar received power and higher as compared to the  $12\times12$~in$^2$ flat square sheet reflector. Moreover, we get a median gain of $20$ dB in case of $24\times24$~in$^2$ and $33\times33$ in$^2$ square sheet reflectors as compared to no reflector. Simulation results also follow a similar trend as in the empirical results. 

\begin{figure}[!t]
	\begin{subfigure}{0.5\textwidth}
	\centering
    \includegraphics[width=0.97\columnwidth]{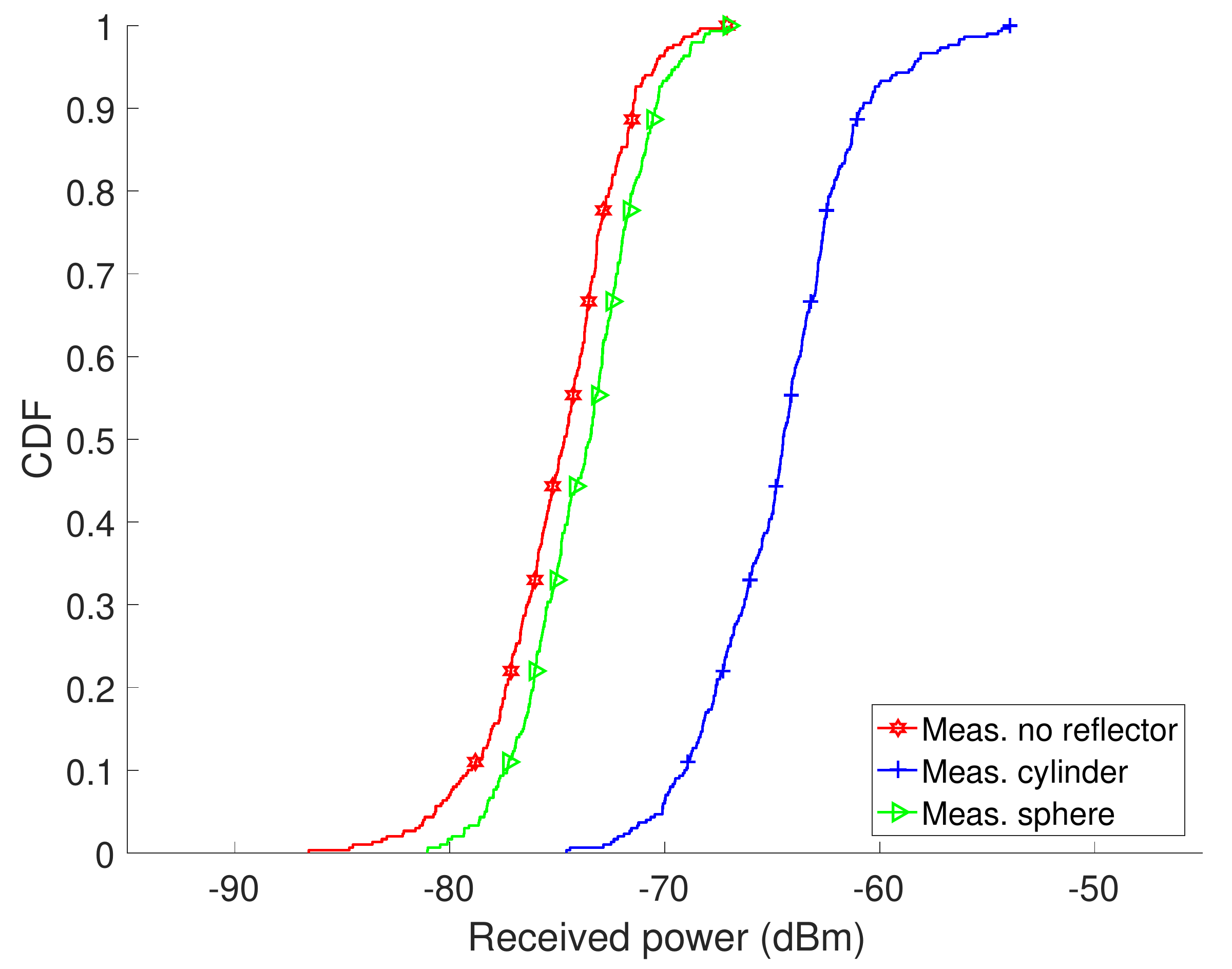}
    \caption{}
    \end{subfigure}	

 	\begin{subfigure}{0.5\textwidth}
    \centering
	\includegraphics[width=0.97\columnwidth]{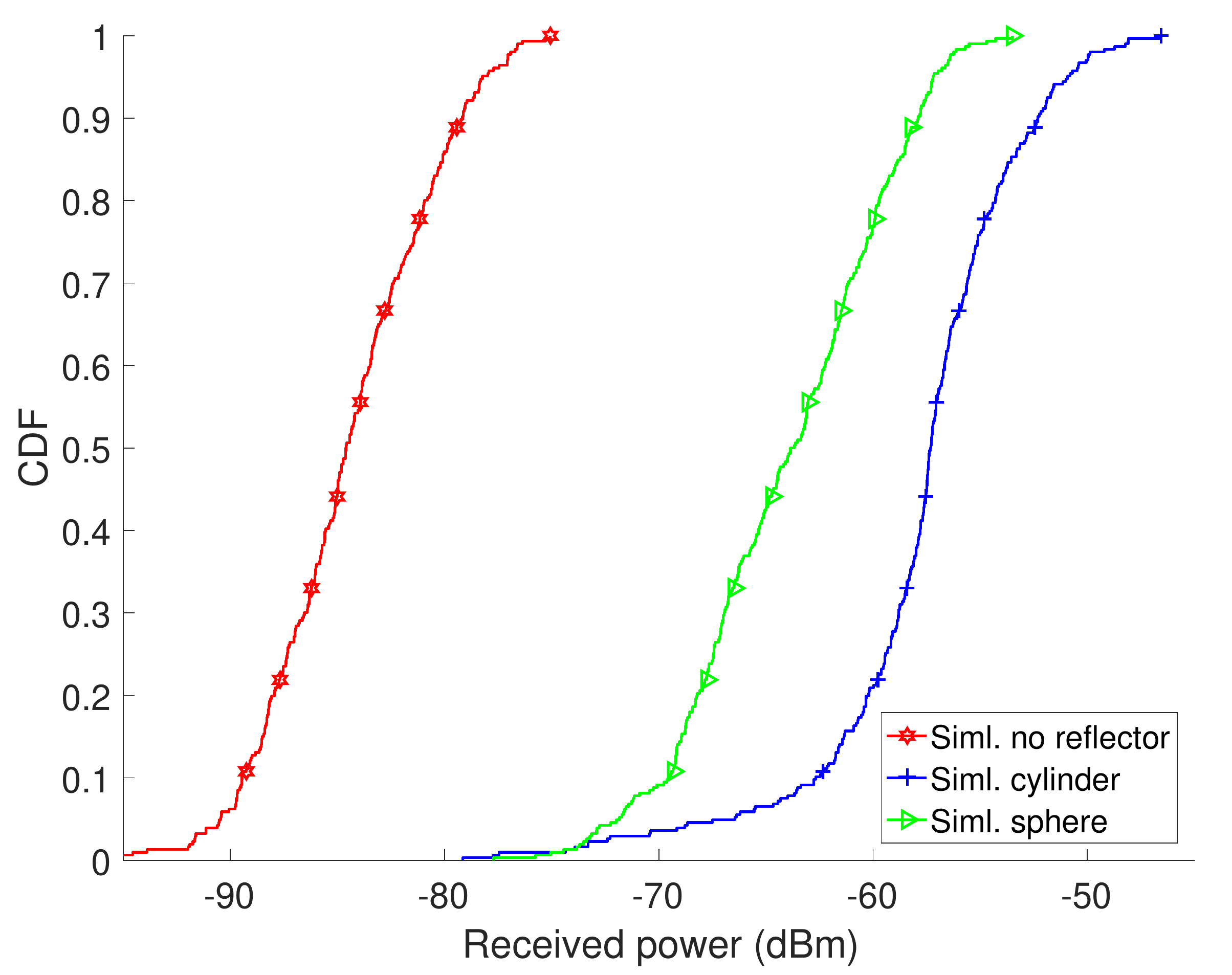}
	  \caption{}  
    \end{subfigure}
    \caption{CDF of received power for cylinder and sphere reflectors obtained using (a) measurements, (b) simulations.}\label{Fig:CDF_Spherical_Cylinder}
\end{figure}

The received power CDFs for measurements with sphere and cylinder reflectors are shown in Fig.~\ref{Fig:CDF_Spherical_Cylinder}. It can be observed that cylindrical reflector has much larger received power as compared to sphere in case of measurements. On the other hand,  ray tracing simulations  show larger received power for the sphere when compared to those that are obtained from measurements. One of the reasons for this behavior can be due to structural construction of the sphere that are used in experiments, where an aluminum sheet is \emph{wrapped} on a spherical mirror ball as compared to other reflectors, and solid aluminum sheets are used rather than thicker metal sheets as in other experiments. We also observe larger variance in the received power in case of simulations as compared to measurements.   


\section{Conclusions}\label{Section:Concluding Remarks}
In this work, channel measurements at $28$~GHz are carried in an NLOS indoor scenario. Passive metallic reflectors of different shapes and sizes are used to enhance the received power, yielding a better signal coverage in the NLOS region. Results show that the flat square sheet reflector provides more favorable coverage in NLOS region  compared to sphere and cylindrical shaped reflectors, where the latter ones scatter the energy more uniformly at shorter distances. For a given rectangular receiver grid, the maximum power is obtained at an azimuth angle of $45^{\circ}$ for flat square sheet reflectors. The measurement results were compared with ray tracing simulations which tend to result in more optimistic coverage. Building on this initial study, our future work includes a more comprehensive measurement campaign in indoor and outdoor environments and developing insights on how to better characterize measurements using ray tracing simulations.  

\bibliographystyle{IEEEtran}





\end{document}